\documentclass[pra,twocolumn,amsmath,amssymb,floatfix,reprint]{revtex4}

\usepackage{graphicx}
\usepackage{color}
\usepackage{pst-node}

\def \ra{{\rightarrow}}
\def \ua{{\uparrow}}
\def \da{{\downarrow}}

\def \cH{{\cal{H}}}
\def \cC{{\cal{C}}}

\def \bA{{\bf A}}

\def \bR{{\bf R}}
\def \br{{\bf r}}

\def \bp{{\bf p}}

\def \b1{{\bf 1}}

\newcommand{\bea}{\begin{eqnarray}}

\newcommand{\eea}{\end{eqnarray}}

\newcommand{\beq}{\begin{equation}}

\newcommand{\eeq}{\end{equation}}

\def \bR{{\bf R}}
\def \br{{\bf r}}

\def \bp{{\bf p}}

\newcommand{\no}{\nonumber\\}
\newcommand{\bra}[1]{\left<#1\right|}
\newcommand{\ket}[1]{\left|#1\right>}
\newcommand{\braket}[2]{\left<#1|#2\right>}

\DeclareMathOperator\sech{sech}

\newcommand{\refe}[1]{Eq.~\ref{#1}}
\newcommand{\reff}[1]{Fig.~\ref{#1}}

\newlength\figurewidth
\newlength\figurefullwidth
\begin{document}

\title{Implementing supersymmetric dynamics in ultracold atom systems}

\begin{abstract}
 Supersymmetric systems derive their properties from conserved supercharges which form a supersymmetric algebra. 
 These systems naturally factorize into two subsystems, which, when considered as individual systems, have essentially the same eigenenergies, and their eigenstates can be mapped onto each other.
 We first propose a one-dimensional ultracold atom setup to realize such a pair of supersymmetric systems. 
  We propose a Mach--Zehnder interference experiment which we demonstrate for this system, and which  can be realized with current technology.
In this interferometer, a single atom wave packet that evolves  in a superposition of the subsystems, gives an
  interference contrast  that is sharply peaked if the subsystems form a supersymmetric pair. 
 Secondly, we propose a two-dimensional setup that implements supersymmetric dynamics in  a synthetic gauge field.  
 \end{abstract}  

\author{M.~Lahrz$^{1,2}$, C.~Weitenberg$^{2}$, L.~Mathey$^{1,2,3}$}

\affiliation{
\mbox{$^{1}$Zentrum f\"ur Optische Quantentechnologien, 
Universit\"at Hamburg, 22761 Hamburg, Germany}\\
\mbox{$^{2}$Institut f\"ur Laserphysik, Universit\"at Hamburg, 22761 Hamburg, Germany}\\
\mbox{$^{3}$The Hamburg Center for Ultrafast Imaging, Luruper Chaussee 149, Hamburg 22761, Germany}
}

\date{\today}

\maketitle


%
Supersymmetry (SUSY) was originally introduced in particle physics beyond the Standard Model but its conceptual structure can be applied outside of high energy theory, giving rise to supersymmetric quantum mechanics~\cite{Witten1981, Sukumar1985c, Cooper1995}.
Here, the algebraic structure of SUSY relates two  Hamiltonians to be SUSY partner Hamiltonians. 
  These share the same eigenspectrum, except for the ground state possibly, and the corresponding eigenstates can be mapped onto each other.
  By mapping a seemingly complicated Hamiltonian on its SUSY partner for which its diagonalization is known, an exact diagonalization can be constructed. 
 This concept has been applied to a wide range of physical problems, e.g. to the hydrogen problem~\cite{Schroedinger1940,Valance1989}, solving the Fokker-Planck equation using imaginary time propagation~\cite{Bernstein1984,Marchesoni1988}, and for multisoliton solutions of the Korteweg-de Vries equation~\cite{Sukumar1986,Fakhri2002}.
 Recent applications have been reported in~\cite{delCampo2014,Ulrich2015}.


%
\begin{figure}[t!]
\includegraphics[width=0.9\figurewidth]{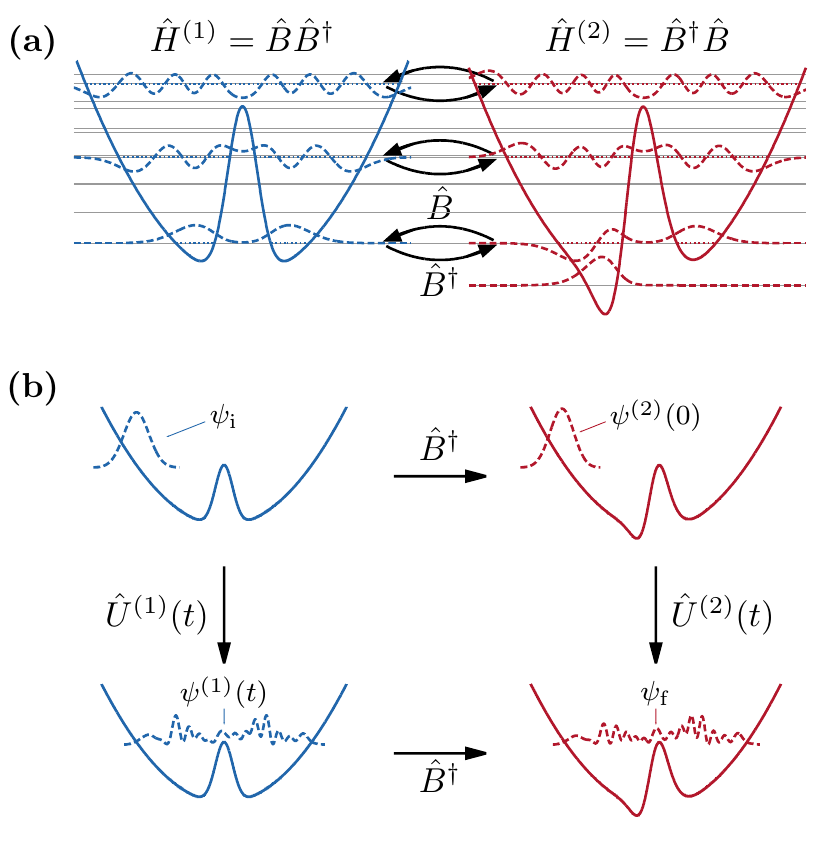}
\caption{%
 Supersymmetric partner systems manifest themselves equivalently as follows.
 (a) The spectra coincide, except for a zero energy state of $\hat{H}^{(2)}$ for unbroken supersymmetry. Additionally, 
 the eigenstates can be mapped onto each other via $\hat{B}^{\dagger}$ and $\hat{B}$.
 (b) The time propagation of each system can be mapped onto each other via $\hat{B}^{\dagger}$  at any time $t$.%
}
\label{fig:cover}
\end{figure}

The simplest supersymmetric algebra consists of a Hamiltonian $\cH$, and a supercharge operator $Q$. We assume that these operate on a $2$-spinor $\psi= (\psi^{(1)}, \psi^{(2)})$, note \footnote{We use the indices $\ua$ and $\da$, and $1$ and $2$ interchangeably.}. The supercharge $Q$ is a conserved quantity of $\cH$, and fulfills the defining equation $\{ Q, Q^{\dagger}\} = \cH$. We choose $Q$ to have the form $Q = B \sigma^{+}$, where $\sigma^{+}$ is the Pauli spin raising matrix, and $B$ is a scalar operator.  With this, $\cH$ takes the form $\cH = \hat{H}^{(1)} \b1_{\ua} + \hat{H}^{(2)} \b1_{\da}$, with $\b1_{\ua/\da} = (\b1\pm\sigma_{z})/2$, and with $\hat{H}^{(1)} = \hat{B} \hat{B}^{\dagger}$ and $\hat{H}^{(2)} = \hat{B}^{\dagger} \hat{B}$. So the system separates into  two scalar subsystems.
 Considering $\hat{H}^{(1)}$ and $\hat{H}^{(2)}$ as individual Hamiltonians, these are  supersymmetric partners. 
  As depicted in \reff{fig:cover}(a), they have the same spectrum, with the possible exception of the ground state of $\hat{H}^{(2)}$, and the operator $\hat{B}$ maps the eigenstates $\psi_{n}^{(2)}$ of 
 $\hat{H}^{(2)}$ onto the eigenstates $\psi_{n}^{(1)}$ of $\hat{H}^{(1)}$. 
  We have $\hat{B}\psi_{n}^{(2)} \sim \psi_{n}^{(1)}$, where $\psi_{n}^{(1)}$ and $\psi_{n}^{(2)}$ have the same eigenenergy $E_{n}$, and similarly $\hat{B}^{\dagger}\psi_{n}^{(1)} \sim \psi_{n}^{(2)}$. 
  The zero energy state of $\hat{H}^{(2)}$ is annihilated by  $\hat{B}$, i.e. $\hat{B}\psi_{0} = 0$, if it exists. This case of unbroken SUSY is indeed realized in the main example  of this paper. We give an example for broken SUSY, in which no such state exists, in App. \ref{app:brokensusy}. 
 The number of zero energy states is the Witten index, and is directly relevant for the proof of the Atiyah-Singer index theorem, Refs. \cite{Cooper1995, Atiyah1973, Berline2004}. 
  Extensions of this supersymmetric algebra use several supercharges $Q_{1}, Q_{2}, \ldots$. 
    This is realized for two-dimensional motion of  a spin-$1/2$ particle in a synthetic gauge field, reviewed in Ref. \cite{dalibardgauge} .
  
 


The supersymmetric relation between the two Hamiltonians translates into the following relation between their time propagators 
\bea\label{eq:operator}
\hat{B}^{\dagger}\hat{U}^{(1)}{\left(t\right)}
&=& \hat{U}^{(2)}{\left(t\right)}\hat{B}^{\dagger}
\eea
where $\hat{U}^{(1,2)}{\left(t\right)} = \exp(-\mathrm{i}\hat{H}^{(1,2)}t/\hbar)$. 
 $\hbar$ is the reduced Planck constant. This statement holds for any time $t$, and is depicted in \reff{fig:cover}(b).

\begin{figure}
\includegraphics[width=0.85\figurewidth]{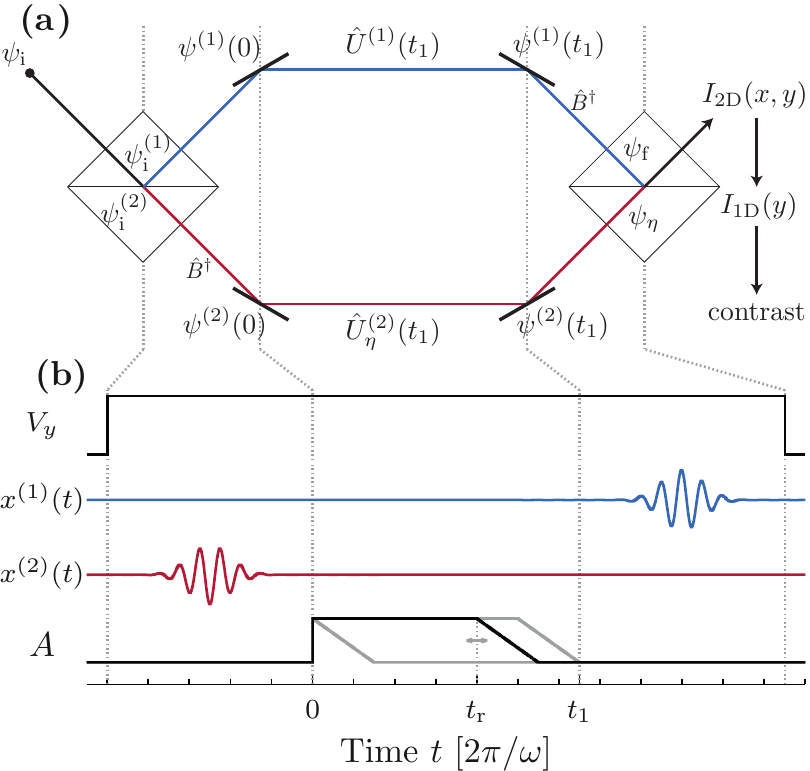}
\caption{%
 (a) Mach--Zehnder interferometer. The initial state $\psi_{\mathrm{i}}$ is coherently split 
to evolve along the two paths corresponding to the supersymmetric partners.
 $\hat{B}^{\dagger}$ is applied before or after the time evolution, depending on the path.
 Finally, the paths are brought to interference, resulting in a contrast measurement.  
 (b) Experimental sequence. 
The beam splitters are realized by a potential barrier  of height $V_{y}$ in $y$-direction, which is smoothly turned on and rapidly turned  off.
 $\hat{B}^{\dagger}$ is approximated by a shaking process.
During this shaking, we turn off the non-harmonic parts of the potentials, i.e. $A = 0$.
The intrinsic time evolution takes place from $t = 0$ to $t_{1}$ including a ramp of $A$.
The time $t_{\mathrm{r}}$ for which $A$ is held constant is varied in the example below.
}
\label{fig:protocol}
\end{figure}

In this paper, we propose to implement and detect the supersymmetric relation of the time propagators in Eq. \ref{eq:operator} in an ultracold atom experiment. 
 In our first example to illustrate our proposal, SUSY is realized in the one-dimensional Schr\"odinger equation, in the second example we discuss atomic motion in a synthetic gauge field in two dimensions. In each case, 
  we consider a single atom that is subjected to an interferometric protocol.
   In the first example,  the atom moves in a one dimensional  harmonic trap potential. 
 The atom is prepared in the harmonic oscillator ground state and then coherently split, either by spatially splitting the trap in one of the confining directions, see e.g. Ref. \cite{Hofferberth2008}, or by applying a $\pi/2$-pulse to a second internal state, such as a hyperfine state.
   Then an experimental approximation of $B^{\dagger}$ is applied to 
  the second  component of the wave packet, see Fig. \ref{fig:protocol}.  
     Both parts of the wave packet are shifted  from the minimum of the harmonic potential by several harmonic oscillator lengths, and for each an additional potential barrier, localized near the harmonic potential minimum is ramped up. The wave packet components then move in each of these potentials. They oscillate in the harmonic potential and scatter off the potential barriers. Then  $B^{\dagger}$ is applied to the first component of the wave packet, and the two components are brought to interference.

This interferometer probes if the two Hamiltonians $\hat{H}^{(1)}$ and $\hat{H}^{(2)}$, with
$\hat{H}^{(i)} = \hat{H}_{kin} + V^{(i)}$, are supersymmetric partners. $\hat{H}_{kin}$ is $\hat{H}_{kin} = - \hbar^{2}/(2 m) \partial_{xx}$, with $m$ being the atom mass. 
 The potentials are $V^{(i)} = V_{osc} + V_{loc}^{(i)}$,  with $V_{osc} = m \omega^{2} x^{2}/2$ where $\omega$ is the trap frequency.  The additional potentials $V_{loc}^{(i)}$ fall off on a length scale $\sigma$. The initial displacement $\bar{x}$ of the wave packets is such that $\bar{x} \gg \max(x_{0}, \sigma)$, where $x_{0}$ is the harmonic oscillator length, $x_{0}=\sqrt{\hbar/(m\omega)}$.
  In this setup  the atom scatters repeatedly off the potential which strongly increases the interference contrast discussed below, which identifies the supersymmetric relation between the two subsystems. We note that other, non-harmonic confining potentials could also be used, see e.g. Apps. \ref{app:brokensusy} and \ref{app:nonharm}.
  As our main example we consider 
\bea
V^{(i)}_{loc}(x) & = & V_{i, s} \exp\Big(-\frac{x^{2}}{2 \sigma^{2}}\Big) +  V_{i, p} x \exp\Big(-\frac{x^{2}}{4 \sigma^{2}}\Big)\label{Vloc}
\eea
Each potential  has a Gaussian term, which we refer to as the $s$-barrier, and a term  of the form $\sim x \exp(-x^{2}/(4 \sigma^{2}))$, which we refer to as the $p$-barrier.
 The $s$-barrier has a width $\sigma$, the $p$-barrier a width $\sqrt{2} \sigma$.
  The $s$-barrier can be experimentally realized with a blue-detuned Gaussian beam, which results in a repulsive potential. The $p$-barrier can be realized by using two Gaussian beams, one blue-detuned and located at $+x_{b}$, with $x_{b}\ll \sigma$, and one red-detuned, located at $-x_{b}$. This combination of two beams realizes  the $p$-barrier potential approximately. 
  Another experimental approach would be to use a spatial light modulator, see e.g. Refs.~\cite{Reicherter1999,Gaunt2012,Nogrette2014,Zupancic2016}.

 Remarkably, in the parameter space of the two  potentials in Eq. \ref{Vloc}, there is a sub-manifold for which  these two systems are  supersymmetric partners. 
To determine this manifold, we consider the  ansatz 
 $\hat{B}^{\dagger} =  (W{\left(x\right)} - \frac{\hbar}{\sqrt{m}}\partial_{x})/\sqrt{2}$. 
 $W{\left(x\right)}$ is the superpotential of the system. Using $\hat{H}^{(1)} = \hat{B}\hat{B}^{\dagger}$ and $\hat{H}^{(2)} = \hat{B}\hat{B}^{\dagger}$, we obtain $\hat{H}^{(i)} = \hat{H}_{kin} + V^{(i)}$, with
 $V^{(i)}{\left(x\right)} = W^{2}(x)/2 \pm \hbar W^{\prime}(x)/(2\sqrt{m})$.
We use
 $W(x) =\sqrt{\hbar\omega}(x/x_{0} + A\exp(-x^{2}/(4\sigma^{2} )))$ for the superpotential. The first term creates the harmonic potential, the second term the barrier potentials. 
 $A$ is the dimensionless amplitude of the Gaussian term.
 This gives
\bea
V^{(i)}(x)
&= &\frac{m \omega^{2} x^{2}}{2}  \mp \frac{\hbar \omega}{2} + \frac{\hbar \omega A^{2}}{2} \exp\Big(-\frac{x^{2}}{2 \sigma^{2}}\Big)\nonumber\\
&& + \frac{2 \hbar \omega A x}{x_{0}} \Big(1 \mp \frac{x_{0}^{2}}{4 \sigma^{2}} \Big) \exp\Big(-\frac{x^{2}}{4 \sigma^{2}}\Big)
\eea
Thus the prefactors of the barrier potentials have to fulfill $V_{1/2, s} = \hbar \omega A/2$ and $V_{1/2, p} = 2 \hbar \omega A (1\mp x_{0}^{2}/(4\sigma^{2}))/x_{0}$.
  We choose $\sigma = x_{0}/2$, resulting in $V_{1, p} = 0$ and $V_{2, p} = 4 \hbar \omega A/x_{0}$. 
  For comparison, we define
\bea\label{eq:Veta}
V_{\eta}(x)
&= &\frac{m \omega^{2} x^{2}}{2}+ \frac{\hbar \omega A^{2}}{2}\exp{\left(-\frac{2x^{2}}{x_{0}^{2}}\right)}\nonumber\\
&&+ \frac{2\eta\hbar \omega A x}{x_{0}}\exp{\left(-\frac{x^{2}}{x_{0}^{2}}\right)}
\eea
as  a family of potentials parametrized by $\eta$. For $\eta=0$, this is $V^{(1)}(x)$, up to an energy offset, and for $\eta=1$, $V_{\eta}(x)$ is $V^{(2)}(x)$.
We will depict the interferometric response between the potential for $\eta=0$ and the potential for an arbitrary $\eta$.
  At $\eta=1$ the response will be  strongly peaked, demonstrating the existence of SUSY.
   In App. \ref{app:Veta}, the spectrum of $V_{\eta}(x)$ is shown as a function of $\eta$. The isospectral feature at $\eta =1$ is clearly visible.



%
\begin{figure}
\includegraphics[width=\figurewidth]{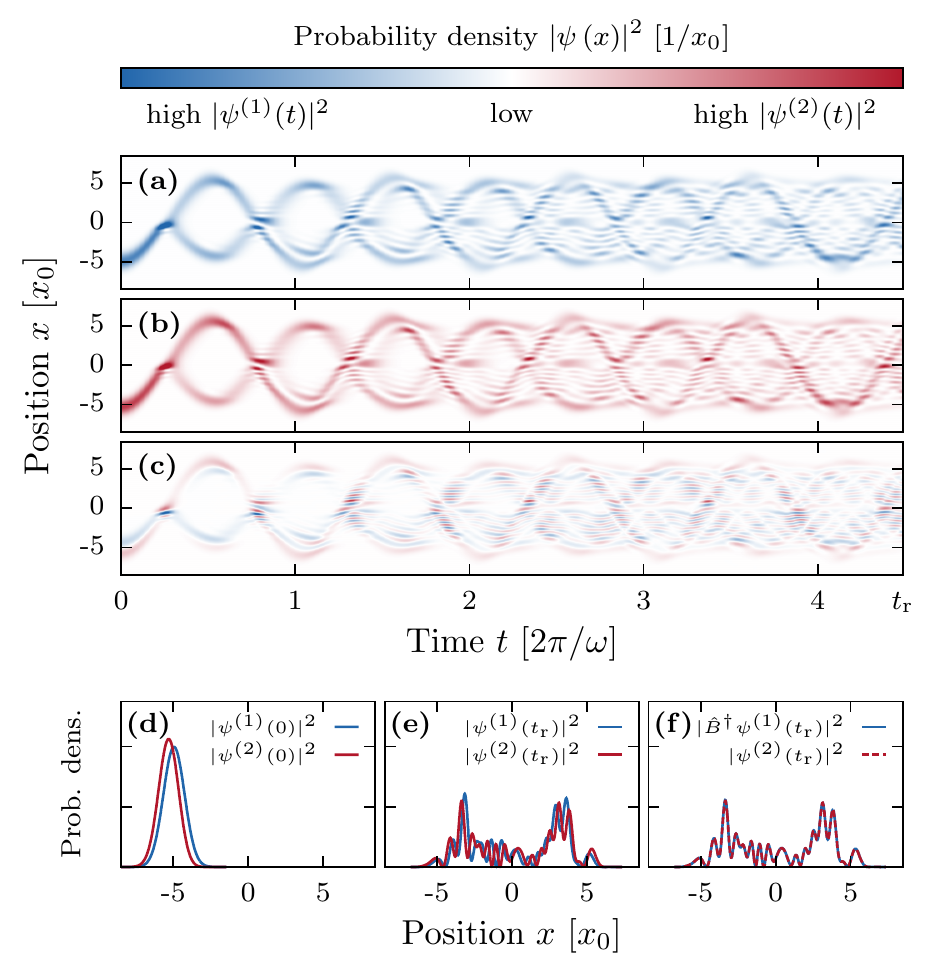}
\caption{
In (a) and (b) we show $|\psi^{(1/2)}(x)|^{2}$, respectively, in (c) the difference of the densities. In (d) we show the initial density in each interferometer path, in (e) the density after the time evolution. In (f) we demonstrate that the two densities coincide after $\hat{B}^{\dagger}$ has been applied to $\psi^{(1)}(x)$.
}
\label{fig:dynamics}
\end{figure}
%


In the protocol, initially the barrier potentials are turned off, $A=0$, see Fig. \ref{fig:protocol} (b).
 The atom is prepared in the ground state,  $\chi_{0}(x) =\left(\pi x_{0}^{2}\right)^{-1/4} \exp(-x^{2}/(2x_{0}^{2}) )$, in one of the internal states. 
 Then a coherent superposition in the interferometer paths is created, by either splitting the trap, or by applying a $\pi/2$-pulse  between internal states.
  Next, we displace the wave function, in the following example  by $\bar{x} = -5\,x_{0}$, and apply $\hat{B}^{\dagger}$ to state $2$. 
   We note that $\hat{B}^{\dagger}$ reduces to the  harmonic oscillator creation operator for $A=0$. 
  When we apply this operator to the shifted ground state wave function, we obtain a superposition of 
  a shifted ground state $\chi_{0}{\left(x\right)}$ and a shifted first excited state $\chi_{1}{\left(x\right)}$ of the harmonic oscillator, that is
 $\zeta_{1}\chi_{1}{\left(x-\bar{x}\right)} + \zeta_{0}\chi_{0}{\left(x-\bar{x}\right)}$.
The amplitudes $\zeta_{0/1}$ depend on  $\bar{x}$ and are  
  $\zeta_{0}^{2} = \bar{x}^{2}/\left(\bar{x}^{2}+2x_{0}^{2}\right)$
and $\zeta_{1}^{2} = 2x_{0}^{2}/\left(\bar{x}^{2}+2x_{0}^{2}\right)$.
For $\bar{x} = -5\,x_{0}$, this gives $\zeta_{0}^{2}  = 49/53$
and
 $\zeta_{1}^{2}  = 4/53$. Thus only a small admixture of the  excited state is necessary to implement  $\hat{B}^{\dagger}$.

To create this admixture, we perform a shaking process of the harmonic trap, i.e.  $V^{(2)}{\left(x,t\right)} = V_{osc}(x-x^{(2)}(t))$, see Fig. \ref{fig:protocol} (b). 
We choose  $x^{(2)}(t) = \delta_{x}\sin(\Omega t) \tau(t)$, where $\delta_{x}$ is the shaking amplitude, $\Omega$ the shaking frequency, resonant with the trap frequency $\Omega=\omega$, and a Gaussian envelope 
 $\tau(t) = \exp(- (t-t_{0})^{2}/(2\sigma_{t}^{2}))$
 with a pulse length $\sigma_{t}$. $t_{0}$ is the pulse time, which we choose to be well separated from turning on the barrier potentials at $t=0$, with $t_{0} = - 5\,\sigma_{t}$.
  This perturbation is of the form $- F(t) \hat{x}$, with $F(t) =m\omega^{2}x^{(2)}(t)$, ignoring an overall energy shift. The $\hat{x}$ operator is $\hat{x} \sim \hat{B}^{\dagger}+\hat{B}$, which is applied to the ground state, giving a small admixture of the first excited state for a small driving term. In App. \ref{app:optshake} we discuss the optimal shaking parameters.  

\begin{figure*}
\includegraphics[width=\figurefullwidth]{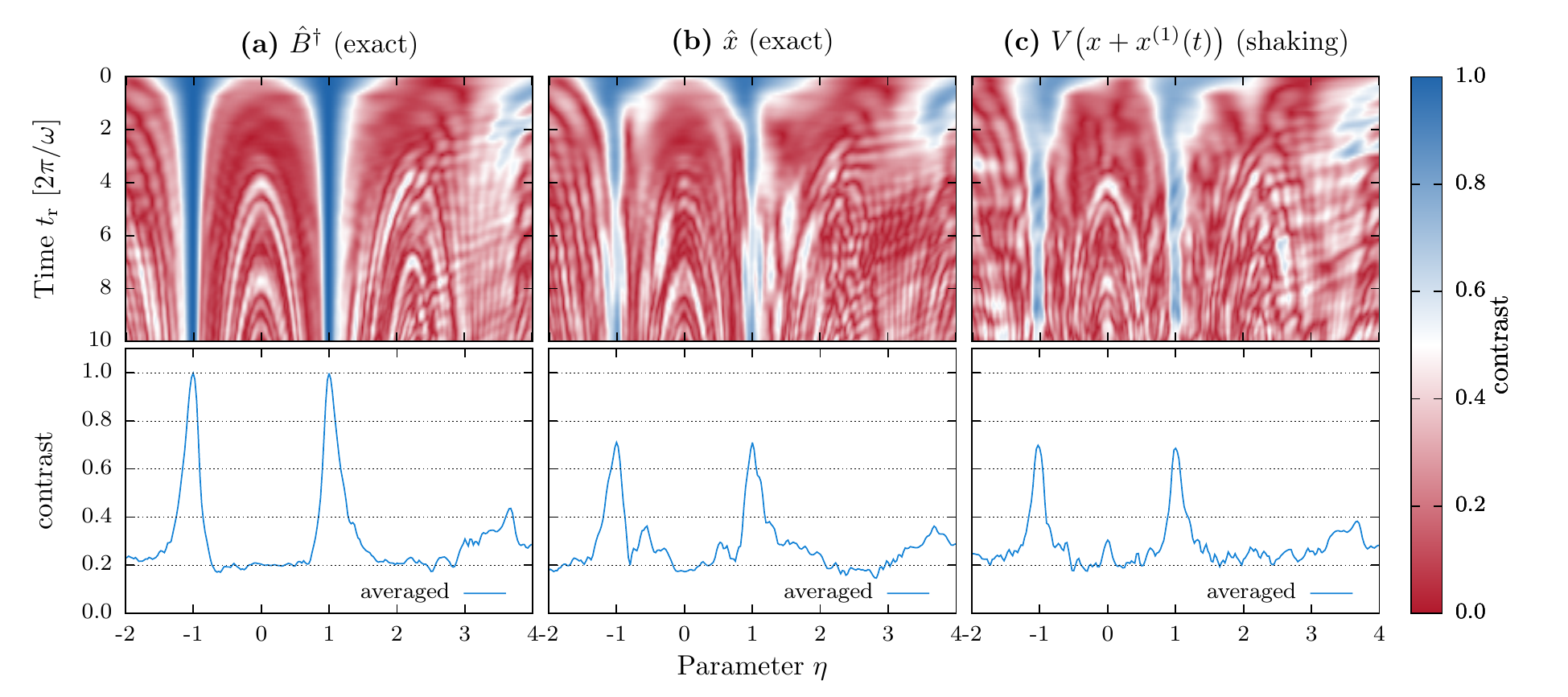}
\caption{
 In the upper panels of (a), (b) and (c) we show the contrast as a function of $t_{r}$ and $\eta$. We interfere $\psi^{(1)}(t_{1})$ with $\hat{B}^{\dagger}  \psi^{(2)}(t_{1})$, $\hat{x}  \psi^{(2)}(t_{1})$ and $\psi^{(2)}(t_{1})$ after the shaking pulse, respectively.  The ramp down time of $A$ is $3\times2\pi/\omega$. 
 In the lower panels, the contrast has been averaged over $t_{r} = 0..10\times2\pi/\omega$.
 The peaks at $\eta = \pm 1$ demonstrate SUSY. 
}
\label{fig:contrast}
\end{figure*}

 Having applied $\hat{B}^{\dagger}$ to state $2$ and  shifted both states by $\bar{x}$, the resulting densities are shown in Fig. \ref{fig:dynamics} (d). The barrier potentials are ramped up at $t=0$.
 We use $A=\sqrt{26}$, see \footnote{The transmission of the barrier approximately $1/2$ for $A = \sqrt{1+\bar{x}^{2}/x_{0}^{2}}$, as we have checked numerically.}.
    The wave packets oscillate in the harmonic potential and scatter off the barrier potentials, see Figs. \ref{fig:dynamics} (a) and (b).  These densities differ at any time, clearly visible by depicting the difference $|\psi^{(1)}(x)|^{2} - |\psi^{(2)}(x)|^{2}$, due to the different barriers  and  initial wave functions.
  The barrier is held constant for a time $t_{r}$, and then ramped down linearly from $t_{r}$ to  $t_{1}$. We show $|\psi^{(1)}(x, t_{r})|^{2}$ and $|\psi^{(2)}(x, t_{r})|^{2}$ in Fig. \ref{fig:dynamics} (e). If we apply $\hat{B}^{\dagger}$ to state $1$,  $| \hat{B}^{\dagger} \psi^{(1)}(x, t_{r})|^{2}$ and $|\psi^{(2)}(x, t_{r})|^{2}$ coincide, as shown in  Fig. \ref{fig:dynamics} (f), which demonstrates supersymmetric dynamics. 
  
  To implement  $\hat{B}^{\dagger}$ after time $t_{1}$ we again perform a shaking pulse, $x^{(1)}(t) = \delta_{x}\sin(\Omega t) \tau(t)$. 
  $\psi^{(1)}(x, t_{1})$ is a superposition of harmonic oscillator states that is controlled by  the initial displacement $\bar{x}$, which results in $\bar{n}\approx \bar{x}^{2}/(2 x_{0}^{2})$, where $\bar{n}$ is the average occupation number. 
  In contrast to the first shaking process, now both  $\hat{B}^{\dagger}$ and  $\hat{B}$ will affect the wave function. Therefore the maximal overlap of the desired and the implemented operator  is approximately $1/\sqrt{2}$. This magnitude can indeed be achieved, in App. \ref{app:optshake} we discuss the optimal parameters. In Fig. \ref{fig:compareoperator} we show the operator that is realized via the shaking process. For the desired energy range of the wave function, this gives satisfactory agreement with  $\hat{B}^{\dagger}$.

 Finally, we bring the interferometric paths to interference. For a realization with two internal states, this is achieved with a $\pi/2$ pulse. For two spatially separated systems, the confining potentials are turned off, and the expanding clouds interfere, see App. \ref{app:interference}. The contrast of these interference patterns is 
 \bea
\cC &=& \frac{2 |\langle \psi_{\mathrm{f}} | \psi_{\eta} \rangle|}{\langle \psi_{\mathrm{f}} | \psi_{\mathrm{f}}\rangle + \langle \psi_{\eta} | \psi_{\eta} \rangle}
 \eea
 where $\psi_{\mathrm{f}}$ and $\psi_{\eta}$ are the states that emerge from the interferometer paths, cf. \reff{fig:protocol}(a). This quantity is depicted in Fig. \ref{fig:contrast} as function of the time $t_{r}$ and of $\eta$, in the upper panels. Here we also compare the contrast for the exact application of $\hat{B}^{\dagger}$, the exact application of $\hat{x}$, and the shaking pulse. The resulting contrast pattern reproduces the exact pattern well. The lower three panels show the contrast integrated over the hold time $t_{r}$. 
The peaks at $\eta=\pm 1$  demonstrate supersymmetry, as desired \footnote{$\eta = -1$ corresponds to replacing $V^{(2)}(x)$ by $V^{(2)}(-x)$, which is also a supersymmetric partner of $V^{(1)}$.}.

{\it Supersymmetry in synthetic gauge fields}. We now apply our protocol to the SUSY algebra with two supercharges $Q_{1}$ and $Q_{2}$, with $\{ Q_{i}, Q_{j}\} = \delta_{ij} \cH$ and $[\cH, Q_{i}]=0$,  with $i,j = 1,2$, and realized in the Pauli equation 
\bea
\cH &=& \frac{(p_{x}+A_{x})^{2}}{2 m} + \frac{(p_{y}+A_{y})^{2}}{2 m} +V(\br) + \frac{\hbar g}{4m} B_{z} \sigma_{z}
\eea
for $V(\br)=0$, and for $g=2$. 
 The supercharges are
 $Q_{1} = ( -(p_{y}+A_{y})\sigma_{x} + (p_{x}+A_{x})\sigma_{y})/(2 \sqrt{m})$
and 
 $Q_{2}= ( (p_{x}+A_{x})\sigma_{x} + (p_{y}+A_{y})\sigma_{y})/(2 \sqrt{m})$, see Refs. \cite{Cooper1995, tomka}. 

  In Refs. \cite{spielman} the synthetic gauge field $A_{x} = B y$ and $A_{y} = 0$, with $B>0$, was proposed and realized.
 This vector potential gives Landau level dynamics in each spin state described by $H   = \omega_{c} (\pi_{x}^{2} + \pi_{y}^{2}- \hbar \sigma_{z})/2 = \hbar \omega_{c} (a^{\dagger} a + (1-\sigma_{z})/2)$, 
 where we introduced $\pi_{x,y} = (p_{x,y} + A_{x,y})/\sqrt{B}$. For this linear vector potential, these are canonical variables, $[\pi_{x}, \pi_{y}]=i \hbar$, 
  so we define $a = (\pi_{x} + i \pi_{y})/\sqrt{2\hbar}$ which are bosonic operators. 
 $\omega_{c} = B/m$ is the cyclotron frequency. 
 The cyclotron motion is centered around $\bR=(X,Y)$, with $X = x + \pi_{y}/\sqrt{B}$ and $Y = y - \pi_{x}/\sqrt{B}$. For $\langle X \rangle = \langle Y \rangle = 0$, the Landau    levels are
 $(a^{\dagger})^{n} \psi_{0} \sim  \exp(i n \phi) \rho^{n}\exp(-\rho^{2}/(4 \rho_{0}^{2}))$
 with $n = 0,1,2,\ldots$, $\rho = \sqrt{x^{2}+y^{2}}$ and the magnetic length
  $r_{0} = \sqrt{\hbar/(m \omega_{c})}$. The ground state $\psi_{0}$ is $\psi_{0} \sim \exp(-\rho^{2}/(4 \rho_{0}^{2}))$
 We note that these are also the eigenstates of the two dimensional harmonic oscillator, $H_{osc} = \bp^{2}/(2m) + m \omega_{c}^{2} \br^{2}/2$, with the same energies $\hbar \omega_{c}(n+1/2)$ and with the angular momentum $l_{z} = n$. 

We propose the following protocol to detect and realize the supersymmetry of this system.
 Initially, the vector potential $\bA$ is turned off, and a harmonic potential $V_{osc}(\br) =m \omega_{c}^{2} \br^{2}/2$ is turned on.
The atom is prepared  in the ground state $\psi_{0}$ in the spin state $\ua$.
 A $\pi/2$ pulse is applied, and then the $a^{\dagger}$ operator, operating on $\da$, is implemented by shaking the harmonic potential
  $V_{osc}(\br-\br_{0}(t))$, while the harmonic potential of $\ua$ is stationary. The circular shaking is given by $\br_{0}(t)=\delta (x \cos(\Omega t), y\sin(\Omega t)) \tau(t) $, with a similar protocol as for the one-dimensional case, generating approximately the transition $n \ra n+1$ and $l_{z} \ra l_{z}+1$. 
  Then the harmonic potential is ramped down, and the gauge field is ramped up.
   Remarkably, any gauge field supports supersymmetry. Therefore, an arbitrary  spatial and temporal $\bA(\br, t)$ can be used on the interferometric paths, as long as the final state is $A_{x} = B y$ and $A_{y} = 0$ again. Then the gauge field is switched off, and the harmonic potential is switched on. The circular shaking process is applied to the state $\ua$, and a $\pi/2$ pulse is used to bring the paths to interference. To detune the system away from supersymmetry, either a potential $V(\br)\neq 0$ can be applied, or $g$ can be tuned as $g= 2 \eta$. Then, a peak at $\eta=\pm 1$ demonstrates supersymmetry.

In conclusion we have demonstrated an interferometric method to realize and detect supersymmetric dynamics in ultra cold atom systems, realizable with current technologies, by laying out two examples. 
 The first example consists of a one-dimensional system which realizes a supersymmetric algebra with one supercharge,
 the second  example consists of two dimensional motion of an atom in a synthetic gauge field, which  conserves two  supercharges.
  We have given a detailed description of the experimental sequence, which includes a beamsplitter step, the application of the supercharge operators, and the constraints on the Hamiltonians of the two subsystems of the supersymmetric system. 
 Using the first realization as an example, we have demonstrated that this protocol gives a sharp interference peak which identifies the system to be supersymmetric.
  From a practical perspective, an intriguing application  
 is a case in which a supersymmetric partner of a desired system is technically easier to realize than the original system, such as the box potential in Fig. \ref{fig:AlternativePotentials}(c). For that case a supersymmetric mapping on the isospectral partner can be implemented.
   More conceptually, the existence of conserved supercharges and the supersymmetric algebraic structure provide a fresh perspective on synthetic gauge fields. Finally, these concepts could potentially be used to test extensions of the Atiyah-Singer index theorem on manifolds with open  boundaries, by creating gauge fields that constraint the motion of a particle to a non-trivial topological subset of the two dimensional plane.

%




We acknowledge support from the Deutsche Forschungsgemeinschaft through the SFB 925 and the Hamburg Centre for Ultrafast Imaging, and from the Landesexzellenzinitiative Hamburg, which is supported by the Joachim Herz Stiftung.



\appendix

\section{Broken and unbroken supersymmetry}\label{app:brokensusy}
For a pair of supersymmetric Hamiltonians with unbroken SUSY, there exists a zero energy  state $\psi_{0}^{(2)}$ in the system described by the Hamiltonian $\hat{H}^{(2)} = \hat{B}^{\dagger}\hat{B}$, which is annihilated by the operator $\hat{B}$, i.e. $\hat{B}\psi_{0}^{(2)} = 0$.
 It can be expressed in terms of the superpotential $W{\left(x\right)}$, that is $\psi_{0}^{(2)} \sim \exp{\left(\sqrt{m}\Omega{\left(x\right)}/\hbar\right)}$ with $\Omega{\left(x\right)} = \int^{x}W{\left(u\right)}\mathrm{d}{u}$.
Since we assume $\psi_{0}^{(2)}{\left(x\to\pm\infty\right)} = 0$ in order to be normalizable, the superpotential must fulfill $W{\left(x\to\pm\infty\right)}=\pm\infty$.
 An example for unbroken SUSY is the one dimensional case presented in this paper, see \reff{fig:cover}, with the zero energy state of $\hat{H}^{(2)}$ visible on the right of panel (a).

\begin{figure}
\includegraphics[width=\figurewidth]{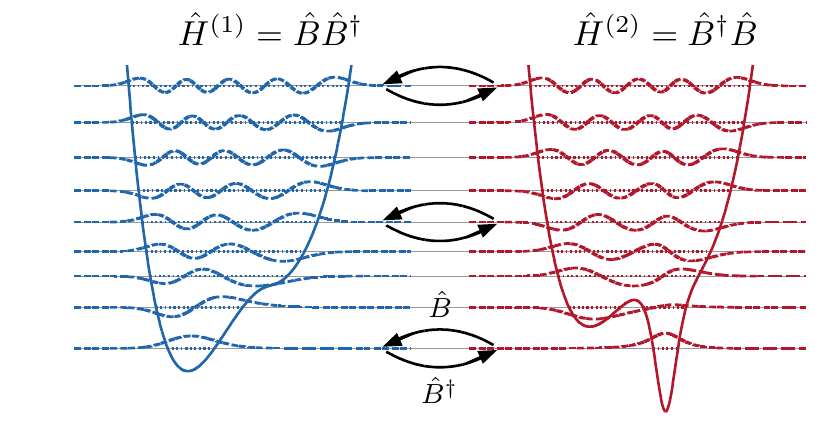}
\caption{Example of a pair of potentials with broken SUSY, see \refe{eq:brokenSUSY}, with $x_{1} = 2$, $x_{2} = 1/2$, and $c = 2$. $\hat{H}^{(2)}$ does not have an additional zero energy state.}
\label{fig:brokenSUSY}
\end{figure}

For supersymmetric Hamiltonians with
 broken SUSY, no zero energy state exists. All eigenstates have a partner state. 
 To give an example for broken SUSY, 
 we choose the superpotential as follows
\begin{align}
W{\left(x\right)} &= \sqrt{2\hbar\omega}\left(\frac{x^{2}}{x_{1}^{2}}+c\tanh{\left(\frac{x}{x_{2}}\right)}\right)\,,
\end{align}
where $x_{1}$ and $x_{2}$ are characteristic length scales and $c$ is a dimensionless constant.
Note, that $W{\left(x\to\pm\infty\right)}=\infty$ which breaks the SUSY.
Further, $W{\left(x\right)}\neq W{\left(-x\right)}$ to avoid the trivial case of $V^{(1)}{\left(x\right)} = V^{(2)}{\left(-x\right)}$.
The SUSY potentials are then given by
\begin{align}\label{eq:brokenSUSY}
V{\left(x\right)}^{(1/2)} &= \hbar\omega \Bigg( \left(\frac{x^{2}}{x_{1}^{2}}+c\tanh{\left(\frac{x}{x_{2}}\right)}\right)^{2}
\no&\quad
\pm\sqrt{\frac{\hbar}{2m\omega}}\left(\frac{2x}{x_{1}^{2}}+\frac{c}{x_{2}}\sech^{2}{\left(\frac{x}{x_{2}}\right)}\right)\Bigg)
\end{align}
where the upper sign is for $V^{(1)}$ and the lower sign for $V^{(2)}$.
We show an example of this family of SUSY potentials in \reff{fig:brokenSUSY}.

For a discussion in periodic potentials, see Ref.~\cite{Khare2004}.


\section{Non-harmonic confining potentials}\label{app:nonharm}
 In our main example for SUSY in a one-dimensional system, we considered a linear term in the  superpotential and thus harmonic confinement in the  potentials $V^{(1)}$ and $V^{(2)}$.
 Here we give two examples in which these potential have a confining potential that is not harmonic, in addition to the example in the previous section.

\begin{figure}
\includegraphics[width=\figurewidth]{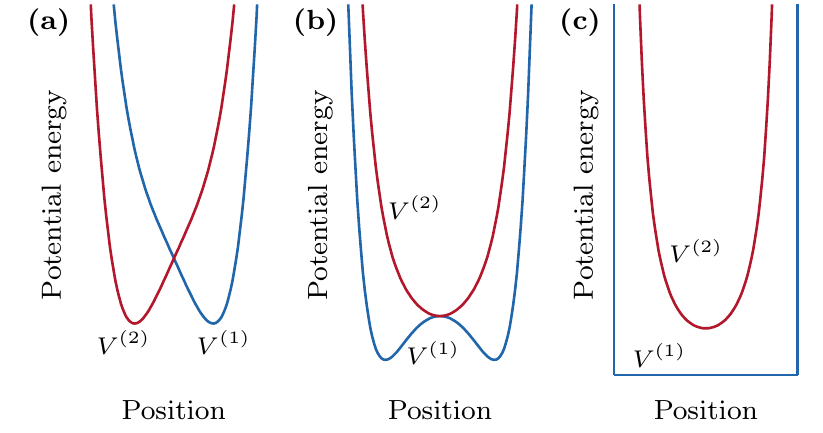}
\caption{Examples of non-harmonic supersymmetric confining potentials. (a) $V^{(1/2)}{\left(x\right)} = 0.01x^{4} \mp 0.2x$. (b) $V^{(1/2)}{\left(x\right)} = 0.01x^{6} \mp 0.3x^{2}$. (c) Box potential $V^{(1)}$ and $V^{(2)} = \pi^{2}\left(2\sin^{-2}{\left(\pi x\right)}-1\right)$.}
\label{fig:AlternativePotentials}
\end{figure}

As a first example, we consider a superpotential of the form $W{\left(x\right)} = cx^{n}$, where $c$ is a constant and $n$ a positive integer.
The potentials are $V^{(1/2)}{\left(x\right)} = c^{2}x^{2n} \mp cnx^{n-1}$ and approach each other asymptotically in the limit $x\to\infty$.
If $n$ is even, the isospectral character of the SUSY partners is trivial, since $V^{(1)}{\left(x\right)} = V^{(2)}{\left(-x\right)}$.
If $n$ is odd, one of the potentials, depending on the sign of $c$, has a double-well structure with a local maximum at $x=0$ while the other has a single minimum at $x=0$.
 We show examples for $n = 2$ and $n = 3$ in \reff{fig:AlternativePotentials}(a+b).

As second example we consider the box potential $V^{(1)}{\left(x\right)} = -\pi^{2}$ for $0 < x < 1$ and infinity else.
The supersymmetric partner is $V^{(2)}{\left(x\right)} = \pi^{2}\left(2\sin^{-2}{\left(\pi x\right)}-1\right)$ for $0 < x < 1$ softening the singularities at the edges of the box, see \reff{fig:AlternativePotentials}(c).
%


\section{Spectrum of the potential  $V_{\eta}(x)$}\label{app:Veta}
\begin{figure}
\includegraphics[width=\figurewidth]{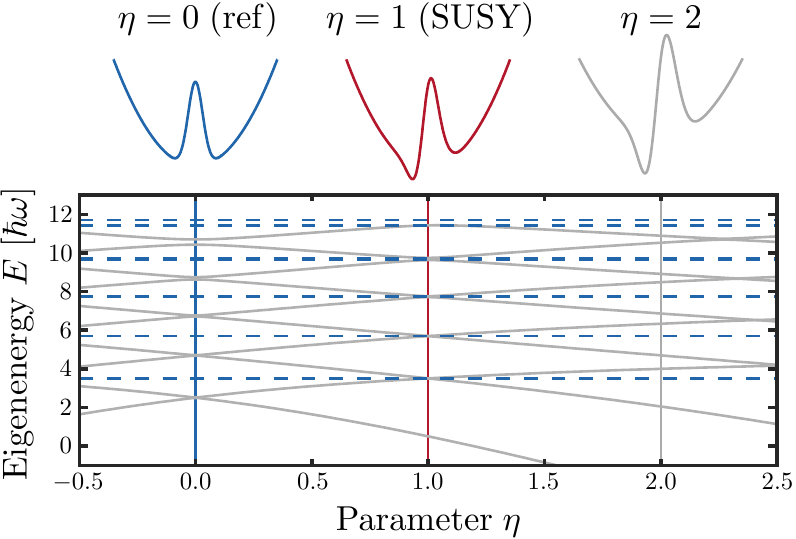}
\caption{
Spectrum of $V_{\eta}{\left(x\right)}$ as a function of the parameter $\eta$ (solid gray) compared to the spectrum of the reference $V^{(1)}{\left(x\right)} - \hbar\omega$ (dashed blue).
The positions of the reference ($\eta = 0$), its SUSY partner ($\eta = 1$) and an arbitrary non-SUSY case ($\eta = 2$) are marked.
The crossing of the constant dashed blue with the gray curves at $\eta=1$ indicates SUSY.
}
\label{fig:spectra}
\end{figure}

We determine the spectra of $V_{\eta}{\left(x\right)}$ as a function of $\eta$, as defined in \refe{eq:Veta}, and compare them to the spectrum of $V_{\eta=0}{\left(x\right)}+\hbar\omega$, see \reff{fig:spectra}.
The spectra of $V_{\eta=1}{\left(x\right)} = V^{(2)}{\left(x\right)}$ and $V_{\eta=0}{\left(x\right)}+\hbar\omega = V^{(1)}{\left(x\right)}$ coincide, except for an additional ground state, indicating the unbroken supersymmetric relation at this point.
%
%
%


\section{Optimal shaking parameters}\label{app:optshake}
We approximate the operator $\hat{B}^{\dagger}$ with a  shaking process, see \reff{fig:protocol}, of the harmonic potential of the form $V{\left(x,t\right)} = V_{\mathrm{osc}}{\left(x+x{\left(t\right)}\right)}$, where $x{\left(t\right)} = \delta_{x}\cos{\left(\Omega t + \phi\right)}\tau{\left(t\right)}$ with the shaking amplitude $\delta_{x}$, the carrier frequency of the shaking $\Omega$, and the phase of the driving $\phi$.
We fix the carrier frequency to $\Omega = \omega$ to be on resonance.
We are free to choose the time envelope $\tau{\left(t\right)}$.
It is useful, although not necessary, to choose a function, which goes to zero at the beginning and at the end of the driving time.
We pick a Gaussian envelope $\tau{\left(t\right)} = \exp{(-(t-t_{0})^{2}/(2\sigma_{t}^{2}))}$ of width $\sigma_{t}$, where $t_{0} = 5\,\sigma_{t}$ is the center of the pulse driving from $t = 0$ to $10\,\sigma_{t}$.

\begin{figure}
\includegraphics[width=\figurewidth]{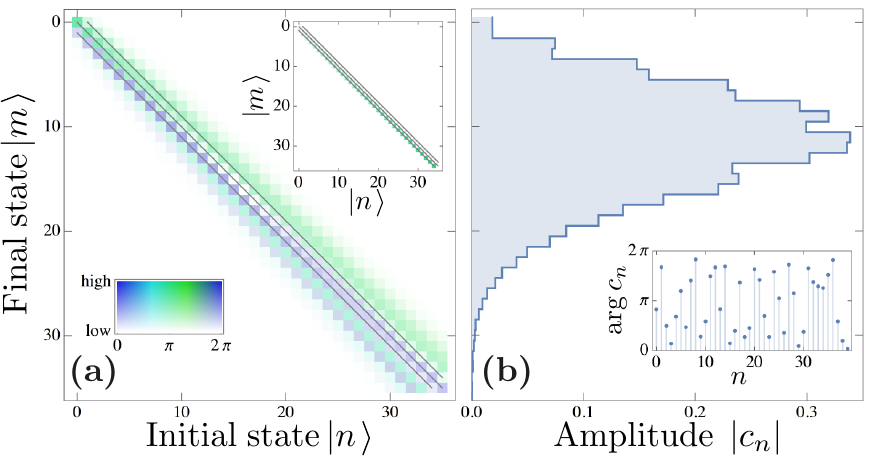}
\caption{(a) Matrix elements in the Fock basis of the shaking process $V{\left(x,t\right)} = V_{\mathrm{osc}}{\left(x+x{\left(t\right)}\right)}$ using the optimal shaking parameters in the first path, $\delta_{x}\sigma_{t} = 0.0470\,x_{0}\times2\pi/\omega$ and $\phi = 0.06\pi$.
Inset: Matrix elements of the harmonic creation operator $\hat{B}^{\dagger} = \sqrt{n+1}\ket{n+1}\bra{n}$.
 (b) Fourier coefficients of a typical desired state, $c_{n} = \braket{n}{\psi^{(1)}{\left(t\right)}}$, in amplitude and phase.
Within the corresponding energy range, $\left|c_{n}\right|$, we find a satisfactory agreement of the two operators shown in (a).
}
\label{fig:compareoperator}
\end{figure}

Based on a path integral ansatz for the forced harmonic oscillator~\cite{Feynman1965}, we have analytic access to the transition matrix elements in the Fock basis, $T_{mn} = \bra{m} \hat{U}(t)\ket{n}$, cf. \reff{fig:compareoperator}~(a). In particular, the transition amplitude  from the ground state $\chi_{0}{\left(x\right)}$ to any state $\chi_{m}{\left(x\right)}$ is given by
\begin{align}\label{eq:Tm0}
T_{m0} &= \frac{T_{00}}{\sqrt{m!}}\left(\mathrm{i} C^{\ast}\right)^{m}
\end{align}
where
\begin{align}\label{eq:Cintegral}
C &= \sqrt{\frac{2}{\pi}}\frac{\omega}{x_{0}}\int_{0}^{T}x^{(2)}{\left(t\right)}\mathrm{d}{t}
\approx -\mathrm{i}\frac{\omega}{x_{0}}\delta_{x}\sigma_{t}\mathrm{e}^{\mathrm{i}\phi}
\end{align}
for $2\omega^{2}\sigma_{t}^{2} \gg 1$ and $t_{0} \gg \sqrt{2}\sigma_{t}$.
Note, that this result only depends on the product $\delta_{x}\sigma_{t}$ and the phase $\phi$.

We apply a shaking process, $x^{(2)}{\left(t\right)}$, in the second path first, starting from the ground state $\chi_{0}{\left(x\right)}$ to create a superposition of the ground state and the first excited state, $\zeta_{0}\chi_{0}{\left(x\right)} + \zeta_{1}\chi_{1}{\left(x\right)}$.
 To create this superposition we demand that
 $T_{00} = \zeta_{0}$, and   $T_{10} \approx \zeta_{1}$.
Since $\zeta_{0}^{2} = \bar{x}^{2}/\left(\bar{x}^{2}+2x_{0}^{2}\right)$, as given in the main text, is close to unity for $\bar{x}/x_{0}\gg1$, we expect the transition probabilities to   higher excited state with $m > 1$ to be negligible.

The shaking process conserves the norm,
$\sum_{m=0}^{\infty}\left|T_{m0}\right|^{2} = 1$,
 so it follows that $\left|C\right| = \sqrt{\ln{\left|T_{00}\right|^{-2}}}$.
Using the assumption $T_{00} = \zeta_{0}$ and \refe{eq:Cintegral}, we find the correct shaking amplitude, that is
\begin{align}
\delta_{x}\sigma_{t} = \sqrt{\ln{\left|\zeta_{0}\right|^{-2}}} = \sqrt{\ln{\left(1+2\frac{x_{0}^{2}}{\bar{x}^{2}}\right)}}\frac{x_{0}}{\omega}\,.
\end{align}
All other matrix elements are now given by \refe{eq:Tm0}, in particular
\begin{align}
T_{10} &= \mathrm{i}T_{00} C^{\ast}
= -\frac{1}{\sqrt{2}}\zeta_{0}\delta_{x}\sigma_{t}\mathrm{e}^{-\mathrm{i}\phi}\,.
\end{align}
Since we aim at $T_{10} = \zeta_{1} \geq 0$ where $\zeta_{1}$ is real, this relation sets the correct amplitude relation to be $\phi = \pi$.
We estimate the error $\epsilon$ of our ansatz by the sum of all higher excitation amplitudes, that is $\epsilon = 1-\left|T_{00}\right|^{2}-\left|T_{10}\right|^{2} \approx 2x_{0}^{4}/\bar{x}^{4}$.

With our choice of $\bar{x} = -5\,x_{0}$ the theoretical amplitude is
$\delta_{x}\sigma_{t} = 0.0442\,x_{0}\times2\pi/\omega$
which is in excellent agreement with our numerical optimized result
$\delta_{x}\sigma_{t} = 0.046\,x_{0}\times2\pi/\omega$.
The phase can be reproduced numerically exact, $\phi = 1.00\pi$.
The error estimate of the ansatz is $\epsilon = 3.2\times10^{-3}$.

In a second step, we apply a shaking process in the first path, $x^{(1)}{\left(t\right)}$, on an unknown state $\psi^{(1)}{\left(t_{1}\right)}$.
Numerically we find the optimal amplitude relation to be $\delta_{x}\sigma_{t} = 0.0470\,x_{0}\times2\pi/\omega$.
Although, this almost agrees with the upper case, we assume that this is coincidental.
The numerical phase is $\phi = 0.06\pi$, while the phase dependency is weakened compared to the upper case.
In \reff{fig:compareoperator} we compare the matrix elements $T_{mn}$ using these optimal shaking parameters with the operator $\hat{B}^{\dagger}$ at an energy range of a typical desired wave function.

%


\section{Interference pattern and contrast}\label{app:interference}
\begin{figure}
\includegraphics[width=\figurewidth]{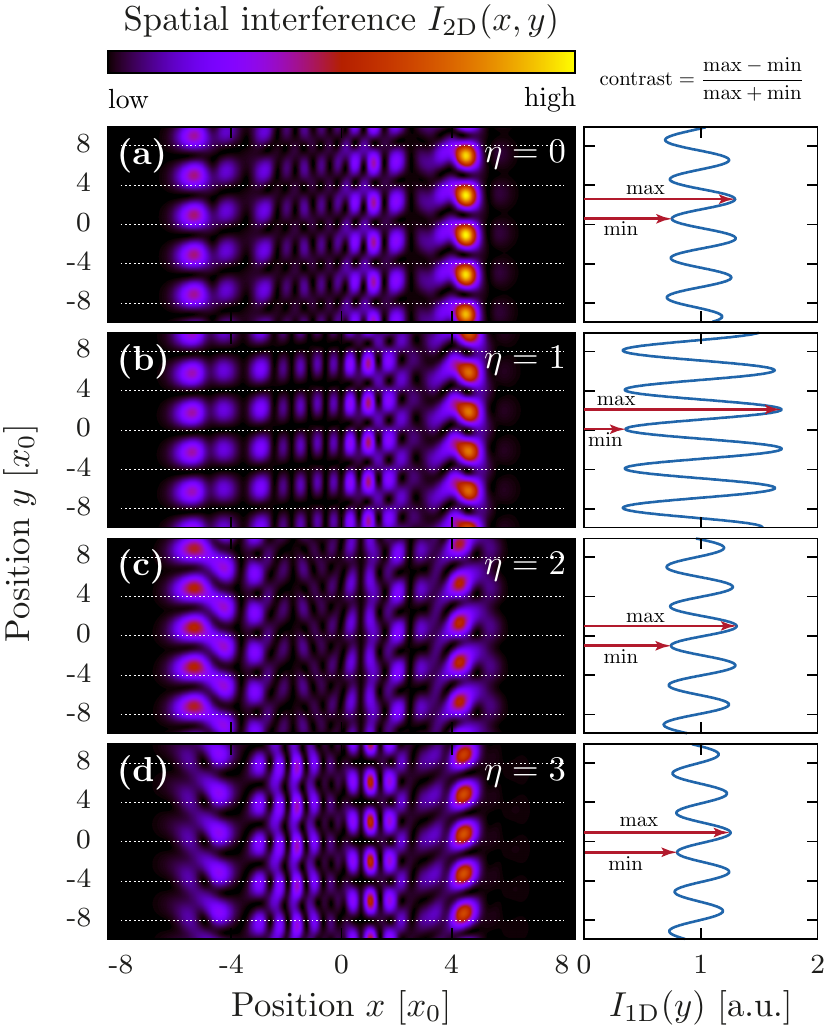}
\caption{
Interference pattern, $I_{\mathrm{2D}}{\left(x,y\right)}$,
 for a holding time $t_{\mathrm{r}} = 4\times2\pi/\omega$ (color plot).
Parallel interference fringes in $x$-direction indicate SUSY in the two paths.
White dashed lines serve as guide to the eye.
Integrating over the $x$-axis gives a one-dimensional interference pattern, $I_{\mathrm{1D}}{\left(y\right)}$, from which we obtain its contrast.
 (a) $\eta = 0$ corresponds to identical potentials in both paths.
 (b) $\eta = 1$ is the SUSY case.
 (c+d) $\eta = 2,3$ show non-SUSY cases.
}
\label{fig:interference}
\end{figure}

For a realization with two spatially separated systems, the confining potentials are turned off to realize the second beam splitter.
The two states, $\psi_{\mathrm{f}}$ and $\psi_{\eta}$, expand rapidly in $y$-direction and form a two-dimensional interference pattern, $I_{\mathrm{2D}}{\left(x,y\right)}$, with interference fringes along the $x$-axis~\cite{Andrews1997,Polkovnikov2006,Hadzibabic2006,Hofferberth2008}, see \reff{fig:interference}.
The envelope of the pattern will depend on $x$ and reflects the spatial density of the states in this direction.
In the case of SUSY, the output states are equal up to a relative phase, $\psi_{\mathrm{f}}{\left(x\right)} \sim \psi_{\eta}{\left(x\right)}$, and the fringes are perfectly parallel to the $x$-axis.

We evaluate the two-dimensional interference pattern $I_{\mathrm{2D}}{\left(x,y\right)}$ by integrating it over the full $x$-axis, that is
\bea
I_{\mathrm{1D}}{\left(y\right)}
&= &\braket{\psi_{\mathrm{f}}}{\psi_{\mathrm{f}}} + \braket{\psi_{\eta}}{\psi_{\eta}}+ 2\left|\braket{\psi_{\mathrm{f}}}{\psi_{\eta}}\right|\cos{\xi{\left(y\right)}}\,,
\eea
where we use the short-hand notation $\braket{\cdot}{\cdot} = \int_{\mathbb{R}}\left|\cdot\right|^{2}\mathrm{d}{x}$.
The phase $\xi{\left(y\right)}$ is the argument of $\braket{\psi_{\mathrm{f}}}{\psi_{\eta}}$ and a linear function in $y$.
The result $I_{\mathrm{1D}}{\left(y\right)}$ is a one-dimensional interference pattern.
For long enough expansion times, we can assume all terms $\braket{\cdot}{\cdot}$ to be independent of $y$.

The contrast $\cC$ of $I_{\mathrm{1D}}{\left(y\right)}$ is defined by the maximum, $\mathrm{max} = \braket{\psi_{\mathrm{f}}}{\psi_{\mathrm{f}}} + \braket{\psi_{\eta}}{\psi_{\eta}} + 2\left|\braket{\psi_{\mathrm{f}}}{\psi_{\eta}}\right|$, and the minimum, $\mathrm{min} = \braket{\psi_{\mathrm{f}}}{\psi_{\mathrm{f}}} + \braket{\psi_{\eta}}{\psi_{\eta}} - 2\left|\braket{\psi_{\mathrm{f}}}{\psi_{\eta}}\right|$, in the following manner:
\bea\label{eq:contrastappendix}
\cC
&= &\frac{\mathrm{max}-\mathrm{min}}{\mathrm{max}+\mathrm{min}}
= \frac{2\left|\braket{\psi_{\mathrm{f}}}{\psi_{\eta}}\right|}{\braket{\psi_{\mathrm{f}}}{\psi_{\mathrm{f}}} + \braket{\psi_{\eta}}{\psi_{\eta}}}\,.
\eea
The contrast is one, if the two states are equal up to a relative phase.
This holds for SUSY partner potentials for any initial state $\psi_{\mathrm{i}}$ and any time $t_{1}$.
Otherwise, the contrast is smaller than one and is determined by fluctuating correlations also in non-SUSY cases resulting in a residual contrast.

In the numerical simulations, we do not consider the expansion in $y$-direction.
Instead, we directly evaluate the second expression of \refe{eq:contrastappendix}.

%


\section{On the displacement $\bar{x}$ }
In our proposal we initialize the system by displacing the ground state of the harmonic confinement by $\bar{x}$.
On the one hand, a large displacement improves the initialization process in the second path of the interferometer, see \reff{fig:protocol}(b).
On the other hand, the correct potential shape, including harmonic confinement and localized central barrier, has to be ensured to energies of the order of $\hbar\omega\left(\bar{x}/x_{0}\right)^{2}/2 \propto \bar{x}^{2}$.
 which might constitute an experimental challenge that limits how large $\bar{x}$ can be chosen.
In this paper, we choose $\bar{x} = -5\,x_{0}$ in our main example, which displays the desired effect.
 We have  considered other values such as $\bar{x} = -3\,x_{0}$.
Here, the contrast after the shaking pulse as depicted in \reff{fig:contrast} (c) gives lower peaks at $\eta = \pm 1$ while the residual contrast away from $\eta = \pm1$ is increased.


\begin{thebibliography}{99}

\bibitem{Witten1981}
E.~Witten,
Nucl. Phys. B \textbf{188}, 513 (1981).

\bibitem{Sukumar1985c}
C.~V.~Sukumar,
J. Phys. A \textbf{18}, 2917 (1985).

\bibitem{Cooper1995}
F.~Cooper, A.~Khare, and U.~Sukhatme,
Phys. Rep. \textbf{251}, 267 (1995).

\bibitem{Schroedinger1940}
E.~Schr\"odinger,
Proc. R.I.A. A \textbf{46}, 9 (1940).

\bibitem{Valance1989}
A.~Valance, T.~J.~Morgan, and H.~Bergeron,
Am. J. Phys. \textbf{58}, 5 (1990).

\bibitem{Bernstein1984}
M.~Bernstein and L.~S.~Brown,
Phys. Rev. Lett. \textbf{52}, 1933 (1984).

\bibitem{Marchesoni1988}
F.~Marchesoni, P.~Sodano, and M.~Zannetti,
Phys. Rev. Lett. \textbf{61}, 1143 (1988).

\bibitem{Sukumar1986}
C.V.~Sukumar,
J. Phys. A \textbf{19}, 2297 (1986).

\bibitem{Fakhri2002}
H. Fakhri, S. Sobhanian, and H. Zahed,
New J. Phys. \textbf{4}, 1 (2002).

\bibitem{delCampo2014}
A.~del~Campo, M.G.~Boshier, and A.~Saxena,
Sci. Rep. \textbf{4}, 5274 (2014).

\bibitem{Khare2004}
A.~Khare and U.~Sukhatme,
J. Phys. A \text{37}, 10037 (2004).

\bibitem{Ulrich2015}
J. Ulrich, D. Otten, and F. Hassler,
Phys. Rev. B \textbf{92}, 245444 (2015).

\bibitem{Infeld1951}
L.~Infeld, T.E.~Hull,
Rev. Mod. Phys. \textbf{23}, 21 (1951).

\bibitem{Atiyah1973}
M.~Atiyah, R. Bott, V.K. Patodi, Invent. Math. \textbf{19}, 279 (1973).


\bibitem{Berline2004}
N. Berline, E. Metzler, M. Vergne,
{\it Heat Kernels and Dirac Operators}, 
 (Springer, Berlin, 2004).


\bibitem{dalibardgauge}
J. Dalibard, F. Gerbier, G. Juzeliunas, and P. \"Ohberg, Rev. Mod. Phys. {\bf 83} 1523 (2011).


\bibitem{Serwane2011}
F. Serwane, G. Z\"urn,T. Lompe, T. B. Ottenstein, A. N. Wenz, and S. Jochim,
Science \textbf{332}, 336 (2011).

\bibitem{Reicherter1999}
M. Reicherter, T. Haist, E. U. Wagemann, and H. J. Tiziani 
Opt. Lett. \textbf{24}, 608 (1999).


\bibitem{Gaunt2012}
A. L. Gaunt and Z. Hadzibabic,
Sci. Rep. \textbf{2}, 712 (2012).

\bibitem{Nogrette2014}
F. Nogrette, H. Labuhn, S. Ravets, D. Barredo, L. Beguin, A. Vernier, T. Lahaye, and A. Browaeys,
 Phys. Rev. X \textbf{4}, 021034 (2014).


\bibitem{Zupancic2016}
P. Zupancic, P. M. Preiss, R. Ma, A. Lukin, M. E. Tai, M. Rispoli, R. Islam, and Markus Greiner,
Opt. Expr. \textbf{24}, 13881 (2016).

\bibitem{Andrews1997}
M.~R.~Andrews, C.~G.~Townsend, H.-J.~Miesner, D.~S.~Durfee, D.~M.~Kurn, and W.~Ketterle
Science \textbf{275}, 637 (1997).



\bibitem{Polkovnikov2006}
A. Polkovnikov, E. Altman, and E. Demler,
PNAS \textbf{103}, 6125 (2006).

\bibitem{Hadzibabic2006}
Z. Hadzibabic, P. Kr\"uger, M. Cheneau, B. Battelier, and J. Dalibard,
Nature \textbf{441}, 1118 (2006).

\bibitem{Hofferberth2008}
S. Hofferberth, I. Lesanovsky, T. Schumm, A. Imambekov, V. Gritsev, E. Demler, and J. Schmiedmayer,
Nature Phys. \textbf{4}, 489 (2008).

%

\bibitem{Feynman1965}
R.P.~Feynman and A.R.~Hibbs,
Quantum mechanics and path integrals
(McGraw-Hill, Inc., London, 1965),
p. 232.


\bibitem{tomka} 
M. Tomka, M. Pletyukhov, V. Gritsev, Scientific Reports {\bf 5} 13097 (2015).



\bibitem{spielman}
I. B. Spielman, Phys. Rev. A {\bf79} 063613 (2009);  Yu-Ju Lin, Rob L. Compton, Karina J. Garcia, James V. Porto, Ian B. Spielman, Nature {\bf 462}, 628 (2009). 




\end{thebibliography}
\end{document}